\documentstyle[12pt,epsf,epsfig,wrapfig]{article}
\def\pole{{\cal P}_e}

\def\polg{{\cal P}_\gamma}
\def\rar{\rightarrow}
\def\lar{\leftarrow}

\begin{document}
\rightline{SLAC-PUB-7319}
\rightline{October 1996}
\rightline{(A/E/I)}
\begin{center}
{\large \bf
The scanning Compton polarimeter for the SLD experiment
\\ }
\vspace{4mm}
{\bf M. Woods}
\\
\vspace{4mm}
{\small Stanford Linear Accelerator Center\\
Stanford University, Stanford, CA 94309\\}
\vspace{3mm}
{\small Representing\\}
\vspace{3mm}
{\bf The SLD Collaboration$^*$\\}
\end{center}

\begin{center}
\vspace{5mm}
\begin{minipage}{130 mm}
\small
{\bf Abstract.}
For the 1994/95 run of the SLD experiment [1] at SLAC, a Compton polarimeter
measured the luminosity-weighted electron beam polarization to be 
$(77.2 \pm 0.5)\%$.  This excellent accuracy is achieved by measuring
the rate asymmetry of Compton-scattered electrons near the kinematic
endpoint.  The polarimeter takes data continuously while the electron and 
positron beams are in collision and achieves a statistical precision of 
better than $1\%$ in a three minute run.  To calibrate the polarimeter
and demonstrate its accuracy, many scans are frequently done.  These include
scans of the laser polarization, the detector position with respect to the
kinematic edge, and the laser power.
\end{minipage}
\end{center}
\centerline{{\it Presented at the Workshop on High Energy Polarimeters}}
\centerline{{\it NIKHEF, Amsterdam, The Netherlands}}
\centerline{{\it September 9, 1996}}

\vskip 3.0in
\small
This work was supported in part by Department of Energy contract
DE-AC03-76SF00515
\pagebreak
 
\normalsize
This polarimeter [2] detects Compton-scattered
electrons from the collision of the longitudinally polarized 45.6 GeV electron
beam [3] with a circularly polarized photon beam. The photon beam is produced
from a pulsed Nd:YAG laser with a wavelength of 532 nm.  After the Compton
Interaction Point (CIP), the electrons pass through a dipole
spectrometer; a nine-channel Cherenkov detector then measures electrons
in the range 17 to 30 GeV.  Figure 1 shows the location of the Cherenkov
detector with respect to the Compton spectrum; the response function
for channel 6 (as determined from an EGS simulation) is indicated as well. 

The counting rates in each Cherenkov channel are
measured for parallel and
anti-parallel combinations of the photon and electron beam helicities.
The asymmetry formed from these rates is given by
$$ A(E)= {R(\rar\rar)-R(\rar\lar) \over R(\rar\rar)+R(\rar\lar)} =
         \pole\polg A_C(E) $$
where $\pole$ is the longitudinal polarization of the electron beam at the CIP,
$\polg$ is the circular polarization of the
laser beam at the CIP, and
$A_C(E)$ is the Compton asymmetry function.

The laser (Spectra Physics GCR130) has a nominal repetition rate of 17 Hz.  It
fires on every 7th electron pulse; the electron pulse rate is 120 Hz. Every
7 seconds the laser fires on the 6th pulse rather than the 7th to avoid any
synchronization of the laser firing with instabilities 
in the electron beam.  Laser off 
pulses are used for determining backgrounds.  The typical Compton collision
rate is approximately 1000 Compton scatters per collision pulse, with 
approximately 100 Compton scattered electrons detected by each of the 7
Cherenkov channels spanning the Compton spectrum.  Typical signal to background
ratio in Channel 7 is about 5:1.

\begin{wrapfigure}{r}{8cm}
\epsfig{figure=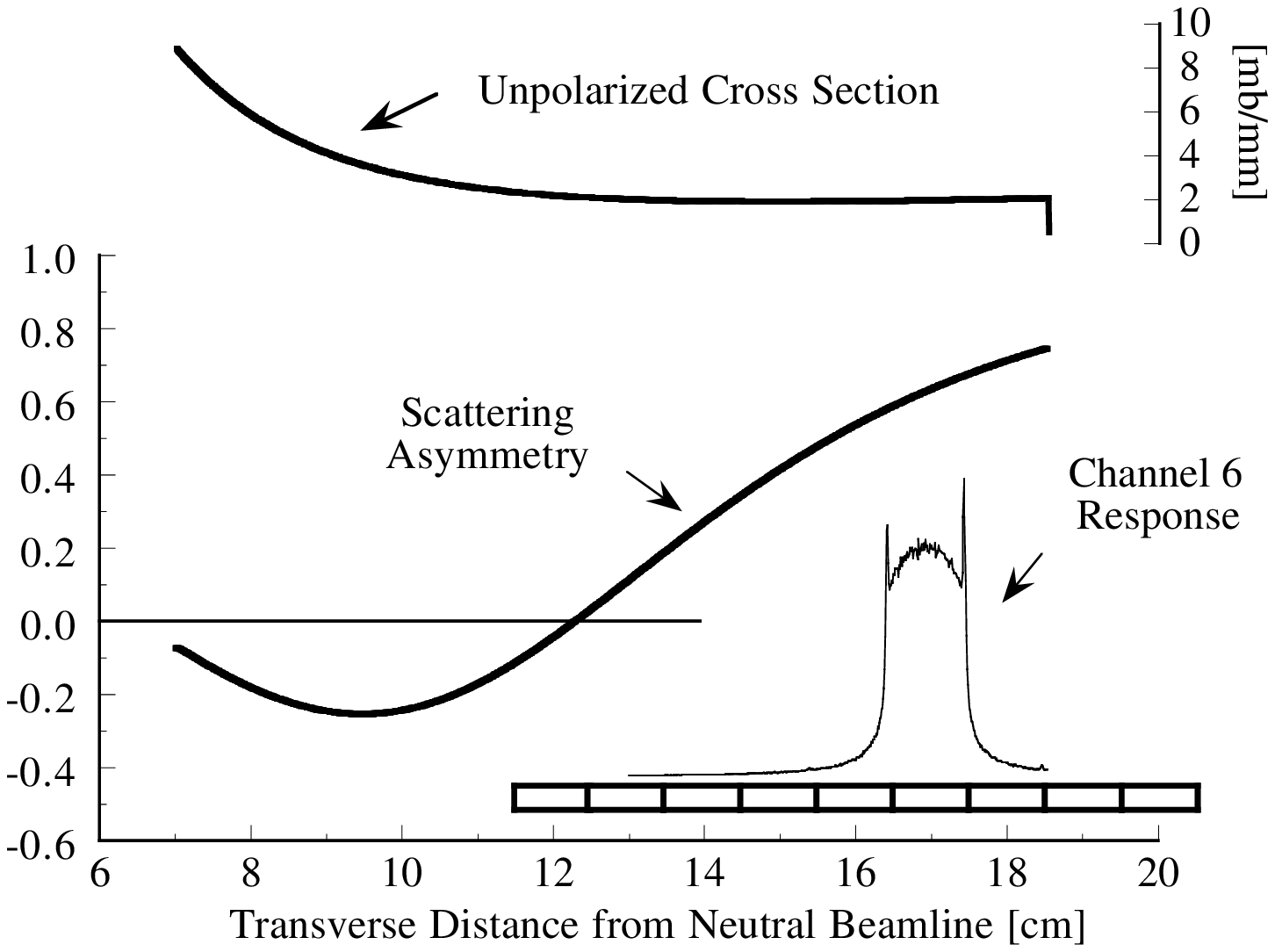,width=8cm}
{\small\bf Figure 1: Compton kinematics}
\end{wrapfigure}

The laser is polarized with
a linear polarizer and two Pockels cells as shown in Figure 2.  The
axes of the linear polarizer and the PS Pockels cell are along the {\it x,y}
axes, while the axes of the CP Pockels cell are along {\it u,v} ({\it u,v}
axes are rotated by $45^{\circ}$ with respect to {\it x,y}).  This 
configuration can generate arbitrary elliptical polarization, and can 
compensate for phase shifts in the laser transport optics. 
Measurements of $\polg$ are made before and after the CIP (see Figure 2).
An harmonic beam sampler (Gentec HBS-532-100-1C-10) transmits $98\%$ of 
the laser power and generates two $1\%$ beams at forward angles of 
$10^{\circ}$,
which preserve the circular polarization, $\polg$, of the main beam to better 
than $0.1\%$.  $\polg$ is determined from photodiode measurements of the
amount of left-polarized and right-polarized light, where the {\it Right} and
{\it Left} photodiodes follow an helicity filter.  The helicity filter is
formed from a quarter waveplate and a calcite prism.  The calcite prism
has different indices of refraction for {\it x} and {\it y} linear polarized
light and splits these components by $5^{\circ}$.  

The {\it Right} ($PD^+$) and {\it Left} ($PD^-$) photodiode signals
and the measured Compton asymmetry in Cherenkov channel 7 ($A_7$), are
well-approximated by the following formulae:
$$ PD^{\pm}={G^{\pm} \over 2}[1 \pm sin({V_{CP}-V_{CP}^T \over 
V_{\lambda/4}^{CP}} \cdot {\pi \over 2}) cos({V_{PS}-V_{PS}^T \over 
V_{\lambda/4}^{PS}} \cdot {\pi \over 2})] $$
$$ A_7 = \pole (A_C^7) [sin({V_{CP}-V_{CP}^{CIP} \over 
V_{\lambda/4}^{CP}} \cdot {\pi \over 2}) cos({V_{PS}-V_{PS}^{CIP} \over 
V_{\lambda/4}^{PS}} \cdot {\pi \over 2})] $$
where $G$ is the photodiode gain; $V_{CP}$ and $V_{PS}$ are the Pockels cell
voltages; 
$V_{\lambda /4}$ is the Pockels cell quarterwave voltage;
$V_{CP}^T$ and $V_{PS}^T$ are the laser transport phase shifts 
to the photodiode diagnostics;
$A_C^7$ is the analyzing power for Cherenkov channel 7;
and $V_{CP}^{CIP}$ and $V_{PS}^{CIP}$ are the laser transport phase shifts 
to the Compton IP.
Measurements of $PD^+$, $PD^-$ and $A_7$ are made at different Pockels cell
voltages ({\it Pockels cell scans}) to monitor the laser transport phase shifts and
the Pockels cell quarterwave voltages.
From these scans
we determined that averaged over the 1994/95 SLD run,
$<\polg> = 99.6 \pm 0.2\%$ (syst) at the CIP.

\begin{figure}
\epsfig{figure=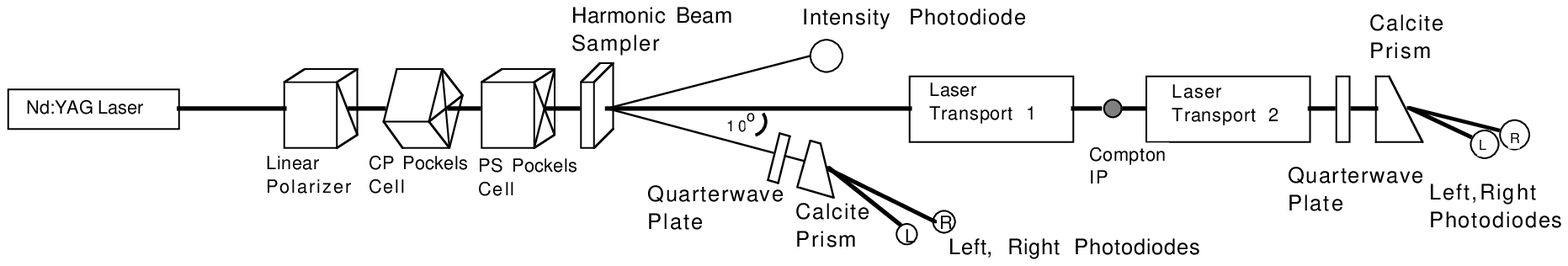,width=15cm}
{\bf Figure 2: Compton laser system}
\end{figure}

The Compton spectrum is characterized (see Figure 1) by a
kinematic edge at 17.4 GeV (180$^{\circ}$ backscatter in the center
of mass frame) where $A_C=0.754$, 
and the zero-asymmetry point at 25.2 GeV (90$^\circ$
backscatter in the center of mass frame).
$A_C(E)$ is modified from the theoretical asymmetry function [4]
by detector resolution effects.  This effect is about 1$\%$
for Cherenkov channel 7 at the Compton edge.
The Compton edge position is accurately determined from Cherenkov detector 
measurements at different detector positions ({\it detector position scans}).
The Compton edge is in channel 7, and we use this
channel to accurately determine $\pole$.  The asymmetry spectrum observed
in channels 1-6 is used as a cross-check.

Polarimeter data are acquired continually during SLC operation and SLD
data logging.
The absolute statistical precision attained in a 3 minute interval
is typically $\delta\pole<1.0\%$. Two-thirds of the polarimeter data are
taken at off-nominal operating conditions (Pockels cell scans, table scans,
laser power scans for linearity tests) for polarimeter calibration and
systematics studies. 
The systematic uncertainties that affect the polarization measurement
are summarized in Table I for the 1994/95 run.  The average 
luminosity-weighted electron beam
polarization at the SLC IP for this run was found to be $<\pole^{IP}>=(77.2
\pm 0.5) \%$.  

\vskip 0.1in
\vbox{
\centerline{{\bf Table I:  Systematic Uncertainties for the SLD
Compton Polarimeter}}
\vskip .05 in
\tabskip=1em plus2em minus.5em
\def\thickfill{\leaders\hrule height1.5pt\hfill}
\vbox{\offinterlineskip
\halign to \hsize
{\vrule width1.5pt#&\strut\hfil#\hfil&\vrule#&
&\strut\hfil#\hfil&\vrule width1.5pt#\cr
 \omit\span\omit\span\omit\span\omit\span\omit
 \thickfill\cr
& Systematic Uncertainty &
& $\delta\pole/\pole$ (\%)  && \cr
 \omit\span\omit\span\omit\span\omit\span\omit
 \hrulefill\cr
& Laser Polarization && 0.20 &&\cr
& Spectrometer Calibration && 0.29&& \cr
& Detector Linearity && 0.50 &&\cr
& Electronics Noise && 0.20 &&\cr
& SLC IP [5] && 0.17 &&\cr
 \omit\span\omit\span\omit\span\omit\span\omit
 \hrulefill\cr
& Total && 0.67&& \cr
 \omit\span\omit\span\omit\span\omit\span\omit
 \thickfill\cr
}}}

\vfill
{\small\begin{description}
\item{[1]}
The SLD Design Report, SLAC-Report-273 (1984).
\item{[2]}
Descriptions of the Compton polarimeter system can be found in the
thesis by R. King, SLAC-Report-452 (1994); and in the thesis by A. Lath,
SLAC-Report-454 (1994).  New additions to the polarimeter not described in
these theses include a higher repetition rate laser, improved laser 
polarization diagnostics, and modification of the spectrometer magnets to 
include a quadrupole.
\item{[3]}
The polarized electron beam for the SLAC Linear Collider
(SLC) is described in a contribution to these proceedings by M. Woods,
SLAC-PUB-7320 (1996).
\item{[4]} 
S.B. Gunst and L.A. Page, 
{\it Phys. Rev.} {\bf 92}, 970 (1953). 
\item{[5]}
There is a small difference between the luminosity-weighted electron beam
polarization relevant for {\it Z} bosons detected by the SLD, and the average
electron beam polarization measured by the Compton polarimeter (see 
reference [3]).
\end{description}}

\pagebreak
\centerline{{\bf $^*$The SLD Collaboration}}



\bigskip
\begin{center}
\bigskip
%
%
%
  \def\iADEL{$^{(1)}$}
  \def\iBOL{$^{(2)}$}
  \def\iBU{$^{(3)}$}
  \def\iBRUN{$^{(4)}$}
  \def\iUCSB{$^{(5)}$}
  \def\iUCSC{$^{(6)}$}
  \def\iCIN{$^{(7)}$}
  \def\iCSU{$^{(8)}$}
  \def\iCOLO{$^{(9)}$}
  \def\iCOL{$^{(10)}$}
  \def\iFER{$^{(11)}$}
  \def\iFRA{$^{(12)}$}
  \def\iILL{$^{(13)}$}
  \def\iLBL{$^{(14)}$}
  \def\iMIT{$^{(15)}$}
  \def\iMASS{$^{(16)}$}
  \def\iMISS{$^{(17)}$}
  \def\iMOSC{$^{(18)}$}
  \def\iNAG{$^{(19)}$}
  \def\iOREG{$^{(20)}$}
  \def\iPAD{$^{(21)}$}
  \def\iPERU{$^{(22)}$}
  \def\iPISA{$^{(23)}$}
  \def\iRUT{$^{(24)}$}
  \def\iRAL{$^{(25)}$}
  \def\iSOGANG{$^{(26)}$}
  \def\iSLAC{$^{(27)}$}
  \def\iTENN{$^{(28)}$}
  \def\iTOH{$^{(29)}$}
  \def\iVAND{$^{(30)}$}
  \def\iWASH{$^{(31)}$}
  \def\iWISC{$^{(32)}$}
  \def\iYALE{$^{(33)}$}
  \def\dead{$^{\dag}$}
  \def\andgen{$^{(a)}$}
  \def\andper{$^{(b)}$}
%
%
$^*$
\mbox{K. Abe                 \unskip,\iNAG}
\mbox{K. Abe                 \unskip,\iTOH}
\mbox{I. Abt                 \unskip,\iILL}
\mbox{T. Akagi               \unskip,\iSLAC}
\mbox{N.J. Allen             \unskip,\iBRUN}
\mbox{W.W. Ash               \unskip,\iSLAC$^\dagger$}
\mbox{D. Aston               \unskip,\iSLAC}
\mbox{K.G. Baird             \unskip,\iRUT}
\mbox{C. Baltay              \unskip,\iYALE}
\mbox{H.R. Band              \unskip,\iWISC
% ...
}
\it
  \vskip \baselineskip                   
  \vskip \baselineskip                   
%
%
%
  \iADEL
     Adelphi University,
     Garden City, New York 11530 \break
  \iBOL
     INFN Sezione di Bologna,
     I-40126 Bologna, Italy \break
  \iBU
     Boston University,
     Boston, Massachusetts 02215 \break
  \iBRUN
     Brunel University,
     Uxbridge, Middlesex UB8 3PH, United Kingdom \break
  \iUCSB
     University of California at Santa Barbara,
     Santa Barbara, California 93106 \break
  \iUCSC
     University of California at Santa Cruz,
     Santa Cruz, California 95064 \break
  \iCIN
     University of Cincinnati,
     Cincinnati, Ohio 45221 \break
  \iCSU
     Colorado State University,
     Fort Collins, Colorado 80523 \break
  \iCOLO
     University of Colorado,
     Boulder, Colorado 80309 \break
  \iCOL
     Columbia University,
     New York, New York 10027 \break
  \iFER
     INFN Sezione di Ferrara and Universit\`a di Ferrara,
     I-44100 Ferrara, Italy \break
  \iFRA
     INFN  Lab. Nazionali di Frascati,
     I-00044 Frascati, Italy \break
  \iILL
     University of Illinois,
     Urbana, Illinois 61801 \break
  \iLBL
     Lawrence Berkeley Laboratory, University of California,
     Berkeley, California 94720 \break
  \iMIT
     Massachusetts Institute of Technology,
     Cambridge, Massachusetts 02139 \break
  \iMASS
     University of Massachusetts,
     Amherst, Massachusetts 01003 \break
  \iMISS
     University of Mississippi,
     University, Mississippi  38677 \break
  \iMOSC
     Moscow State University,
     Institute of Nuclear Physics,
     119899 Moscow,
     Russia    \break
  \iNAG
     Nagoya University,
     Chikusa-ku, Nagoya 464 Japan  \break
  \iOREG
     University of Oregon,
     Eugene, Oregon 97403 \break
  \iPAD
     INFN Sezione di Padova and Universit\`a di Padova,
     I-35100 Padova, Italy \break
  \iPERU
     INFN Sezione di Perugia and Universit\`a di Perugia,
     I-06100 Perugia, Italy \break
  \iPISA
     INFN Sezione di Pisa and Universit\`a di Pisa,
     I-56100 Pisa, Italy \break
  \iRUT
     Rutgers University,
     Piscataway, New Jersey 08855 \break
  \iRAL
     Rutherford Appleton Laboratory,
     Chilton, Didcot, Oxon OX11 0QX United Kingdom \break
  \iSOGANG
     Sogang University,
     Seoul, Korea \break
  \iSLAC
     Stanford Linear Accelerator Center, Stanford University,
     Stanford, California 94309 \break
  \iTENN
     University of Tennessee,
     Knoxville, Tennessee 37996 \break
  \iTOH
     Tohoku University,
     Sendai 980 Japan \break
  \iVAND
     Vanderbilt University,
     Nashville, Tennessee 37235 \break
  \iWASH
     University of Washington,
     Seattle, Washington 98195 \break
  \iWISC
     University of Wisconsin,
     Madison, Wisconsin 53706 \break
  \iYALE
     Yale University,
     New Haven, Connecticut 06511 \break
  \dead
     Deceased \break
  \andgen
     Also at the Universit\`a di Genova \break
  \andper
     Also at the Universit\`a di Perugia \break
\rm
%

\end{center}

\end{document}